\documentclass{optica-article}

\journal{opticajournal} 
\usepackage{ulem}

\articletype{Research Article}
\usepackage{siunitx}
\usepackage{array}

\usepackage{float}
\usepackage{gensymb} 
\usepackage{makecell}
\usepackage{color,soul}
\usepackage{xcolor}
\definecolor{delectricblue}{RGB}{102, 178, 255}
\colorlet{lightdelectricblue}{delectricblue!30}

\newcommand{\comment}[1]{}

\begin{document}
\title{Frequency and intensity noise of a grating-tuned external-cavity quantum cascade laser}

\author{Irene La Penna\authormark{1,2,*}, Tecla Gabbrielli\authormark{1,2}, Cristina Rimoldi\authormark{1,2}, Davide Mazzotti\authormark{1,2}, Jérôme Faist\authormark{3}, Luigi Consolino\authormark{1,2}, Simone Borri\authormark{1,2}, Paolo De Natale\authormark{1,2} and Francesco Cappelli\authormark{1,2}}

\address{\authormark{1}LENS - European Laboratory for Non-Linear Spectroscopy, Via Carrara 1 - 50019 Sesto Fiorentino FI, Italy\\
\authormark{2}CNR-INO - Istituto Nazionale di Ottica, Largo Fermi, 6 - 50125 Firenze FI, Italy\\
\authormark{3}Institute for Quantum Electronics, ETH Zürich, Zürich, Switzerland\\}

\email{\authormark{*}lapenna@lens.unifi.it} 


\begin{abstract*} 
Quantum cascade lasers (QCLs) are semiconductor-heterostructure devices known for their emission in the mid-infrared and THz spectral regions. Due to their operating regime, their intrinsic linewidth is significantly narrower compared to bipolar semiconductor lasers. Here, we demonstrate that by implementing an external-cavity (EC) configuration based on a commercial diffraction grating, we have successfully induced a Fabry-Pérot QCL to emit on a single mode with a broadly-tunable wavelength in the range 4.29--\SI{4.44}{\micro \meter}. This very simple setup enhances the laser’s performance in terms of threshold current and emitted power. We further prove that the EC configuration positively impacts the laser’s noise properties. In particular, the intrinsic linewidth is substantially reduced, the full linewidth is also decreased (depending on the integration timescale), and the relative intensity noise is slightly reduced. These characteristics, which hold within the whole tuning range, make the EC-QCL a good candidate for spectroscopy applications where broad tunability and narrow linewidth are highly demanded.
\end{abstract*}

\section{Introduction}
Quantum cascade lasers (QCLs) are semiconductor-heterostructure-based devices emitting mid-infrared and THz radiation~\cite{capasso_quantum_1994,faist_quantum_2013} and are well-known for their narrow intrinsic linewidth~\cite{bartalini_observing_2010}. This appealing characteristic, which is correlated to a high coherence, is primarily due to the unipolar nature of the device relying on intersubband operation. The fact that the lasing transition is not based on electron-hole recombination, reduces carrier density fluctuations and the linewidth-enhancement factor, also known as alpha-factor~\cite{henry_theory_1982,bartalini_observing_2010, cappelli_intrinsic_2015}, suppressing the conversion of intensity noise to frequency noise. 
Regarding the full linewidth, even simple Fabry-Pérot (FP) waveguide QCLs have a narrow linewidth with respect to diode lasers (10--100-MHz-range~\cite{fleming_fundamental_1981,coldren_relative_2012}). Owing to their emission frequency, lying in the mid-infrared (MIR) and THz spectral regions, QCLs have been extensively and successfully used for various spectroscopic applications~\cite{wysocki_widely_2005, wysocki_widely_2008, hancock_direct_2009, young_external_2009, gambetta_mid-infrared_2011, weidmann_atmospheric_2011, galli_absolute_2013, lambrecht_broadband_2014, galli_mid-infrared_2014, bartalini_frequency-comb-assisted_2014, nadeem_sensitive_2018, gambetta_versatile_2018, borri_high-precision_2019, consolino_quantum_2020}, 
where broad tunability has been obtained, among the others, through coupling of the QCL to an external cavity~\cite{luo_grating-tuned_2001, maulini_external_2006, mohan_room-temperature_2007, xu_tunable_2007, wittmann_heterogeneous_2008, hugi_external_2009, tsai_external-cavity_2010, meng_broadly_2015, matsuoka_external-cavity_2018, gu_tunable_2021, bayrakli_external_2022}. 

In this work, we report about the frequency noise characterization of a homemade external-cavity (EC) quantum cascade laser, made up of a FP QCL with a diffraction grating in Littrow configuration. This simple and cost-efficient tabletop setup allows us to turn a multimode QCL into a broadly-tunable, single-mode laser, with enhanced performance in terms of emitted power, frequency and intensity stability. Previous works demonstrated that the presence of optical feedback does not degrade the performance of MIR and THz QCLs in terms of mode hopping and intensity noise~\cite{mezzapesa_intrinsic_2013, zhao_relative_2019}. A similar methodology was used in ref.~\cite{zhao_strong_2020} to investigate the evolution of frequency noise, full and intrinsic linewidth of MIR QCLs under optical feedback from a flat mirror with different coupling strengths. In this work, we aim at characterizing a particular type of external cavity with a fixed coupling strength and a feedback selective in wavelength, focusing in particular on the effect within the whole laser's tuning range. For this purpose, the noise analysis has been performed by employing an unbalanced interferometer as frequency-to-amplitude converter. This setup was specifically adopted due to its intrinsic broadband operation capability as opposed to e.g. using molecular absorption lines~\cite{consolino_qcl-based_2019}. This type of characterization is particularly worthwhile in view of using the laser source under test for spectroscopy and sensing applications, often requiring both tunable and stable emitters.  

\section{Methods and discussion}
\begin{figure}[htb!]
\centering\includegraphics[width=\columnwidth, keepaspectratio]{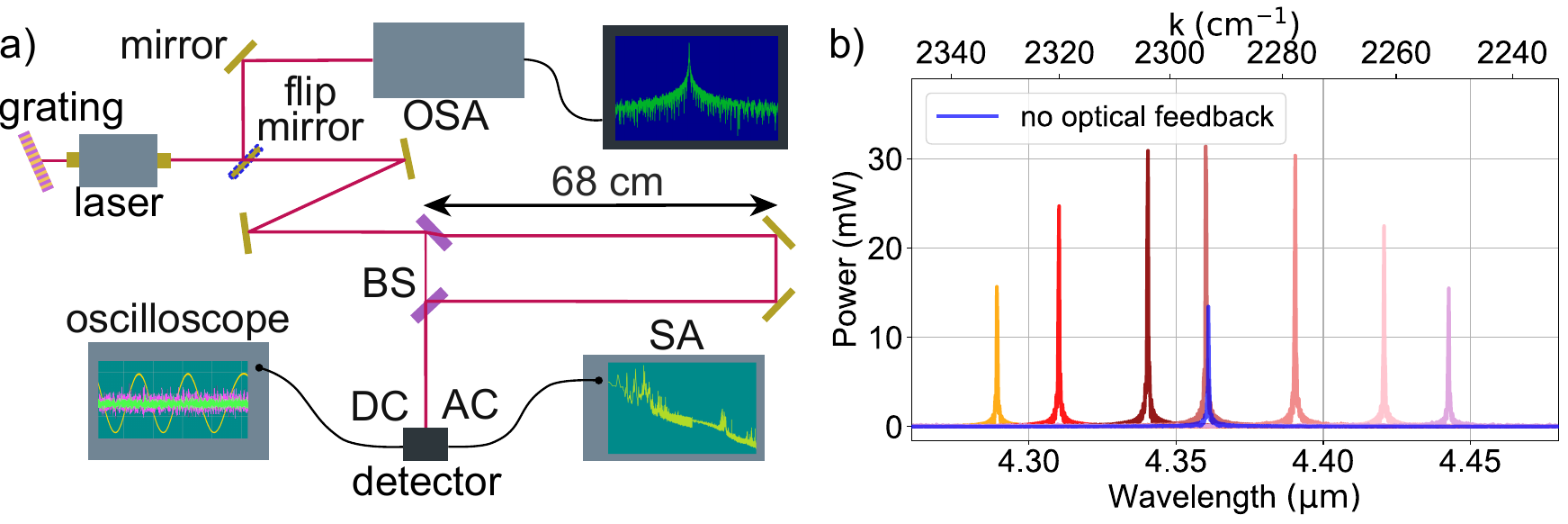}
\caption{(a) Experimental setup for measuring the frequency noise of the EC-QCL. The laser's output beam is sent to an unbalanced Mach-Zehnder interferometer. The interference between the two time-shifted beams coming from the two unequal arms results in intensity fluctuations acquired via the real-time spectrum analyzer (SA). These are converted into frequency noise power spectral density (FNPSD) via the conversion factor found by means of the oscilloscope. By blocking one arm, the intensity noise can be measured. This is done for all of the modes depicted in Fig.~\ref{fig:setup}(b), whose spectrum is monitored via an optical spectrum analyzer (OSA). (BS: beam splitter, DC: detector's DC signal, AC: detector's AC signal)\\ 
(b) Optical spectra of the laser operated at a bias current of 430 mA in free-running (blue trace) and EC configuration (other traces). 
Each peak actually represents the power of the respective mode. The blue trace represents the free-running laser's emission, centered at approximately \SI{4.36}{\micro m} and with a power of \SI{13.46}{mW}. When the grating vertical angle is optimized for the central mode, the power reaches \SI{31.5}{mW}. By adjusting the grating's tilt angle, the laser emission wavelength is tuned accordingly. 
With a careful optimization of the vertical angle, a broad tuning range of the emission is obtained, covering \SI{153.6}{nm} or, equivalently, \SI{80.6}{cm^{-1}}, with a modal power ranging from 15.5 to 31.5~mW, as represented by the peaks in this figure.}
\label{fig:setup}
\end{figure}

Broad tuning of QCLs can be obtained in a variety of configurations, all of which basically rely on the injection of optical feedback in the waveguide by means of an external cavity~\cite{hugi_single-mode_2013}. In case of a QCL with two emitting facets, the most simple and effective way to obtain an EC laser is through the back-extraction Littrow setup, consisting of a diffraction grating placed in front of the back-emitting facet, feeding back into the laser the $-1$ diffraction order. This is the configuration adopted in our work (see Fig.~\ref{fig:setup}(a)), using a commercial diffraction grating from Thorlabs (model GR1325-30035). 
The grating has an efficiency of 23\% at a wavelength of \SI{4.36}{\micro m} for the vertical polarization, the laser beam is collimated at the back extraction facet and the length of the external cavity is 9.0~cm. No external attenuator is placed within the EC. We remark here that the external cavity length is much shorter than the coherence length of the QCL emission, since for a laser full linewidth of $\sim 1$~MHz it exceeds 100~m. Therefore, the provided feedback is coherent with the intrawaveguide field. The amount of feedback back-coupled to the laser facet (limited by the mode matching) is estimated to be $\sim 85\%$. The QCL is a standard FP ridge-waveguide device with two emitting clived facets (with no coating). The waveguide length is 4.2~mm. Additional details of the QCL are given in section~A of Supplement~1. When no optical feedback is present, the lasing threshold is at a bias current of 400~mA at a temperature of \SI{-8}{\celsius}. The laser emits on a single mode centered at \SI{4.36}{\micro m} from threshold up to 440~mA. Above 440~mA of driving current the emission turns multimode (details of the LIV curves and optical spectra are in section~B of Supplement~1). The optical feedback induced by the grating has several effects, the most evident being an increase in the power emitted from the front facet, increasing by a factor 2.3 in the best condition (see Fig~\ref{fig:setup}(b)). A further effect is a reduction of the threshold current, from 400~mA without feedback to 360~mA with the EC. The feedback provided by the grating forces the laser to emit in single mode at any operating point of the LIV curve. The whole characterization was performed with a constant bias current of 430~mA and a constant temperature of \SI{-8}{\celsius}. 
In this condition, the tuning range of the EC-QCL, obtained by rotating the grating, is \SI{153.6}{\nano m}, or equivalently \SI{80.6}{\cm^{-1}}. This tuning range is significantly broader than the spectral coverage corresponding to the multimodal operation of the free-running QCL (see Fig.~S2(a) in section~B of Supplement~1). With the EC, the emitted power spans from \SI{15.5}{mW} at the edges to \SI{31.5}{mW} at the center of the tuning range. 
The power emitted in the EC configuration is always higher than the power emitted by the free-running laser, see Fig.~\ref{fig:setup}(b).

\subsection{Measurement procedure and data analysis}

One of the main aims of the characterization performed in this work is to investigate how the external cavity impacts the laser noise. Initially, we focus on frequency noise, which can be measured with several methods~\cite{borri_high-precision_2019}. We choose to use an unbalanced Mach-Zehnder interferometer as a frequency-to-intensity converter. The Mach-Zehnder configuration was preferred to the Michelson one due to the drastically reduced amount of unwanted feedback sent back to the laser. The scheme is depicted in Fig.~\ref{fig:setup}(a). We scan the laser frequency by modulating the bias current and measure the typical sinusoidal interference fringes at the output of the interferometer with a dedicated detector. The fringes are then used to perform the intensity-to-frequency reconversion of the measured noise. The peak-to-peak voltage of the fringes, $\mathrm{V_{pp}}$, multiplied by $\mathrm{\pi \cdot \tau}$, with $\mathrm{\tau}$ being the time delay between the two arms of the interferometer, gives the conversion factor $\mathrm{\pi \tau V_{pp}}$ from frequency to intensity fluctuations. A more comprehensive computation of the frequency-to-intensity conversion factor is given in section~C of Supplement~1. This conversion factor is used to transform the intensity noise spectra acquired with a real-time spectrum analyzer (SA) in frequency noise power spectral density (FNPSD) spectra. As explained in detail in section~C of Supplement~1, we acquire the frequency fluctuations when the DC signal is at half of the interference fringe, where the frequency-to-intensity conversion is the highest. For this reason, the lower bound of the interferometer to distinguish frequency from intensity noise is given by the intensity noise power spectral density (INPSD) at half fringe. By blocking one arm, we let half of the light intensity arrive on the detector, and we measured the corresponding intensity noise. The INPSD and FNPSD spectra are recorded for the free-running laser and for the EC-tuned modes depicted in Fig.~\ref{fig:setup}(b), while monitoring the spectra with an optical spectrum analyzer (OSA) (see Fig.~\ref{fig:setup}(a)). The plots of FNPSD, INPSD, and the detector thermal background, presented in Fig.~S3 of section~E in Supplement~1, show that frequency fluctuations can be distinguished from intensity fluctuations up to a Fourier Frequency of 70~MHz, corresponding to the bandwidth of the interferometer. 

The FNPSD spectra reveal several laser characteristics. Of utmost importance is the laser spectral lineshape, i.e., the power spectral distribution $I_\mathrm{E}(\nu)$ of the field. This can be derived from the FNPSD $S_{\nu}(f)$ through following equation~\cite{fermigier_frequency_1998,stephan_laser_2005}:
\begin{equation}
    I_{\mathrm{E}}(\nu) = E_0^2 \int_{0}^{\infty} \mathrm{cos} \bigl[ 2 \pi (\nu - \nu_0) \tau \bigr] \cdot \mathrm{exp} \left[ -4 \int_{f_{\mathrm{start}}}^{\infty} S_{\nu}(f) \frac{\mathrm{sin}^2(\pi f \tau)}{f^2} df \right] d\tau
\label{eq:optical PSD}
\end{equation}
where $E_0$ is the mean amplitude of the laser's optical field (neglecting amplitude fluctuations), $\mathrm{\nu}$ represents the optical frequencies, and $f$ stands for the Fourier frequencies in the noise spectra. Here, $S_{\nu}(f)$ is the single-sided Fourier transform of the laser instantaneous frequency. The FNPSD typically has a polynomial form with two main components, a white noise component $h_{\mathrm{0}}$ and a flicker-noise component $h_{\mathrm{-1}/f}$. 
The white noise is an intrinsic property of the laser related to spontaneous emission. By identifying this white intrinsic noise, we can calculate through Eq.~\ref{eq:optical PSD} the optical Power Spectral Density (PSD) to be a Lorentzian lineshape, while its full width at half maximum (FWHM) is the so-called intrinsic linewidth, also known as the fundamental laser linewidth or Schawlow-Townes linewidth~\cite{henry_theory_1982,pollnau_spectral_2020}. In real-life lasers, this white noise is buried in a certain amount of 1/f-noise, arising from the electrical noise coming from the driving electronics and/or from carrier dynamics in the active medium~\cite{yamanishi_electrical_2014}. The linewidth of the resulting spectrum is drastically increased and can be calculated numerically by using Eq.~\ref{eq:optical PSD} to find the lineshape and then computing its FWHM. The use of Eq.~\ref{eq:optical PSD} implies a double integration over the FNPSD. In particular, for the first integration over the Fourier frequencies, the starting frequency $f_\mathrm{start}$ defines the chosen range and the so-called integration time, which should always be specified when a laser linewidth is given. On the other hand, for the intrinsic linewidth $\mathrm{\delta \nu}$ there is a much simpler formula arising from the analytical computation of the Lorentzian lineshape:
\begin{equation}
    \delta \nu = \pi h_0
    \label{eq:intrinsic_lw}
\end{equation}
with $\mathrm{h_0}$ being the white noise level of the FNPSD, usually found at high Fourier frequencies (i.e. short timescales). According to the Schawlow-Townes-Henry theory~\cite{henry_theory_1982}, the expected intrinsic linewidth of a semiconductor laser can be computed using the following equation: 
\begin{equation}
    \delta \nu = \frac{h \nu}{P}\frac{c^2}{4 \pi n_\mathrm{g}^2} (\alpha_{\mathrm{i}}+\alpha_{\mathrm{m}}) \alpha_{\mathrm{m}} n_\mathrm{sp} (1 + \alpha_{\mathrm{E}}^2)
    \label{eq:intrinsic_lw_th}
\end{equation}
where $h$ is the Planck constant, $\nu$ is the laser emission frequency, $P$ is the power emitted by the laser, $c$ is the speed of light, $n_\mathrm{g}$ is the group refractive index, $\alpha_{\mathrm{i}}$ are the internal (waveguide) losses, $\alpha_{\mathrm{m}}$ are the mirror losses, $n_\mathrm{sp}$ is the spontaneous-emission factor or population inversion parameter, and $\alpha_{\mathrm{E}}$ is Henry linewidth-enhancement factor.

\subsection{Discussion}
\begin{figure}[htb!]
\centering\includegraphics[width=\columnwidth, keepaspectratio]{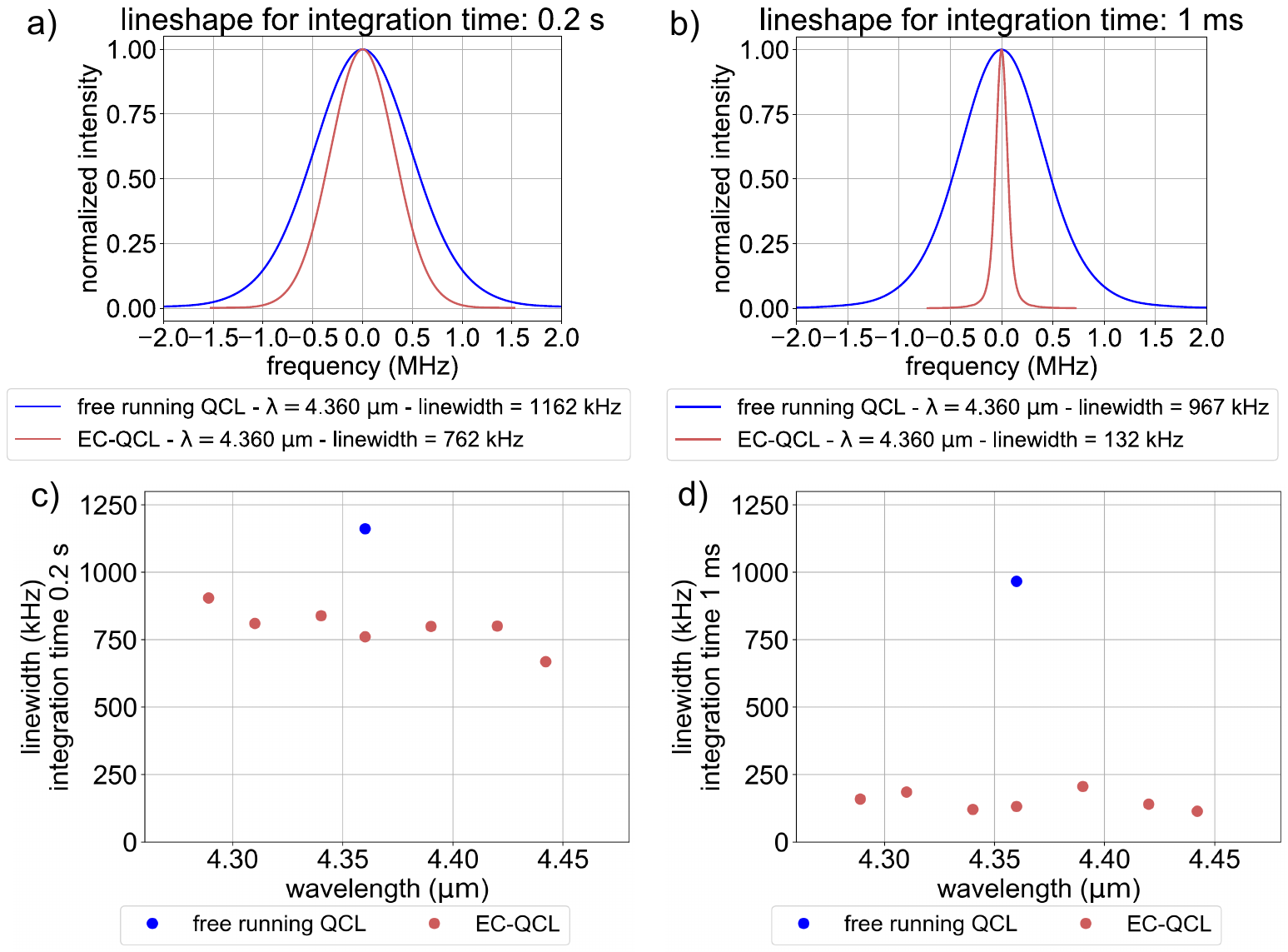}
\caption{(a) and (b) show the lineshape of the free-running QCL (blue) and the one of the EC-QCL emitting at \SI{4.36}{\micro m} (red), computed numerically via Eq.~\ref{eq:optical PSD} using different starting integration frequencies $f_\mathrm{start}$: 5 Hz in (a), corresponding to an integration time of 0.2 s, and 1 kHz in (b), corresponding to an integration time of 1 ms. The linewidth of the free-running together with all EC modes is plotted against their respective wavelength in (c) for the case of an integration time of 0.2 s and in (d) for integration time 1 ms. While varying the integration time does not change substantially the free-running QCL linewidth, which stays around 1 MHz, it has a significant impact on linewidth of the EC-QCL modes which result between 100 and 200 kHz.}
\label{fig:lw_4plots}
\end{figure}

The interferometer has a cutoff frequency $f_\mathrm{c}$ of approximately 70~MHz, due to the delay time of 4.53~ns related to the unbalancing between its two arms. The FNPSD spectra have been, therefore, multiplied by a compensating factor $(1+(f/f_\mathrm{c})^2)^2$ accounting for this cutoff to flatten the spectra at high frequencies. 
\begin{figure}[htb!]
\centering\includegraphics[width=\columnwidth, keepaspectratio]{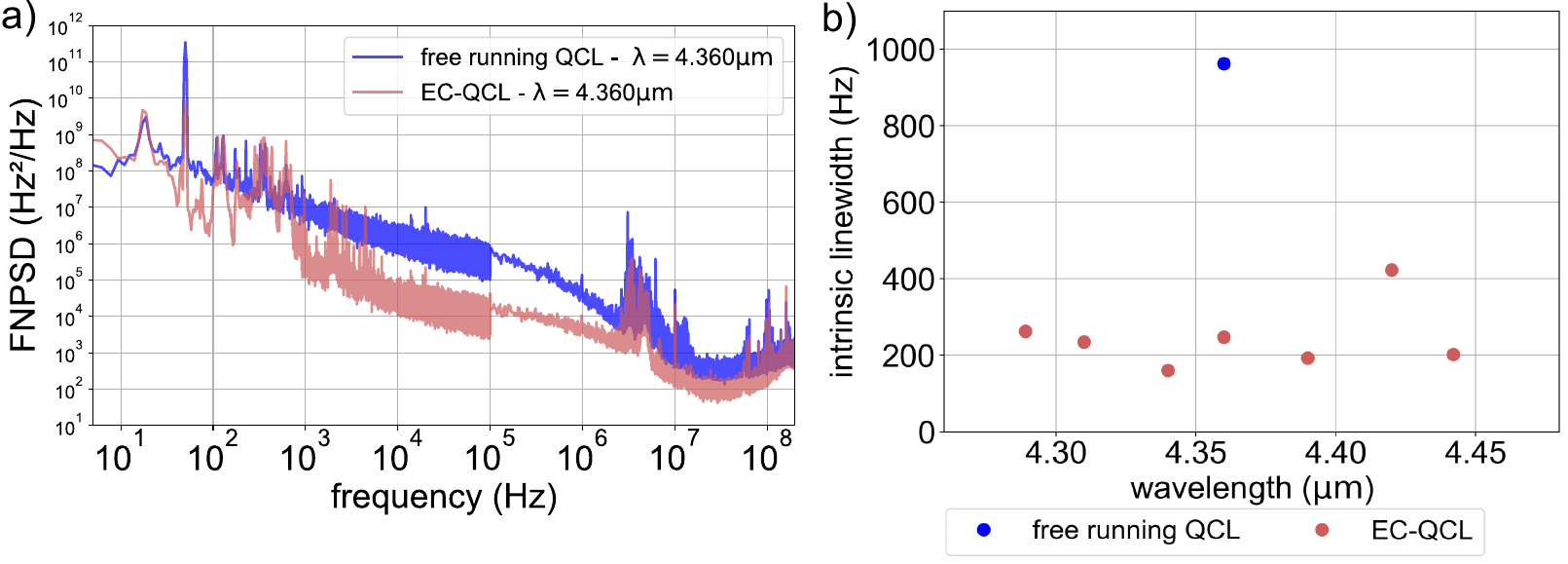}
\caption{(a) FNPSD spectra of the free-running QCL (blue) and of the EC-QCL emitting at \SI{4.36}{\micro m} (red). Here, the spectra are calculated as the single-sided Fourier Transform of the laser instantaneous frequency. The single spectrum is obtained as a superposition of two acquisitions according to the different Fourier frequency range. The FNPSD of the EC-QCL is generally lower for any mode of the tuning range. The peaks at low frequencies (50~Hz--5~kHz) are due to the mechanical vibrations of the grating. The spectra show an unphysical growth after 70~MHz, which is an artifact given by the compensating factor introduced to account for the interferometer's cutoff. The whole set of FNPSD, including the related background spectra, can be found in section~E of Supplement~1.  \\
(b) Intrinsic linewidth of the QCL as a function of the wavelength. The intrinsic linewidth is computed via Eq.~\ref{eq:intrinsic_lw}, where $h\mathrm{_0}$ is the level of white noise measured by averaging the spectra in Fig.~\ref{fig:FN-intrinsic-lw}(a) in the frequency range 30--50~MHz. The intrinsic linewidth of the EC-QCL is significantly narrower than the one of the free-running QCL, for any of the wavelengths.}
\label{fig:FN-intrinsic-lw}
\end{figure}
Fig.~\ref{fig:FN-intrinsic-lw}(a) shows an example of the FNPSDs acquired for the free-running QCL (blue trace) and for the EC-QCL (red trace), when both operate at \SI{4.36}{\micro m}. A complete characterization was performed at other wavelengths in the tuning range of the EC-QCL. The related spectra are shown in Fig.~S4 of Section~E in Supplement~1. 
Starting from the FNPSD and using Eq.~\ref{eq:optical PSD}, it is possible to retrieve the lineshape of the QCL using different integration times, as shown in Fig.~\ref{fig:lw_4plots}(a) and \ref{fig:lw_4plots}(b). The low-frequency peaks present in all of the FNPSD spectra of the EC-QCL (see Fig.~\ref{fig:FN-intrinsic-lw} or Fig.~S4 of Supplement~1), which originate from mechanical vibrations of the grating, have an important impact when calculating the spectral lineshape and eventually the linewidth. In particular, this is evident when the starting integration frequency $f_\mathrm{start}$ is smaller than 1 kHz, as shown in Fig.~\ref{fig:lw_4plots}(a) and \ref{fig:lw_4plots}(b), where two values of $f_\mathrm{start}$ has been used (5~Hz and 1~kHz), resulting in an integration time of 0.2 s and 1 ms for the lineshapes of the free-running QCL (blue line) and the EC-QCL emitting at \SI{4.36}{\micro m} (red line). Despite the mechanical noise contribution, the full linewidth (i.e. the FWHM of the lineshape) of the EC-QCL is narrower than the one of the free-running QCL in the whole tuning range, and even for long integration times as summarized in Fig.~\ref{fig:lw_4plots}(c) and \ref{fig:lw_4plots}(d). 

The acquired FNPSDs are used to extract via  Eq.~\ref{eq:intrinsic_lw} the corresponding intrinsic linewidths, where $h\mathrm{_0}$ is the level of white noise measured by averaging the FNPSD spectra in the frequency range 30–50 MHz. This analysis leads to the results shown in Fig.~\ref{fig:FN-intrinsic-lw}(b). The EC-QCL linewidths are significantly smaller than the intrinsic linewidth of the free-running laser of a factor $\sim 5$. This is expected due to the higher ratio of stimulated over spontaneous emission, the main factor affecting the intrinsic linewidth~\cite{bartalini_observing_2010,cappelli_intrinsic_2015}. In particular, at $\lambda = \SI{4.36}{\micro \meter}$ the intrinisc linewith is $\delta \nu_\mathrm{fr} = \SI{960}{\hertz} $ and $\delta \nu_\mathrm{EC} =  \SI{250}{\hertz} $, and the power is $P_\mathrm{fr} = \SI{13.5}{\milli \watt} $ and $P_\mathrm{EC} = \SI{31.5}{\milli \watt} $ for the free-running and the EC QCL, respectively. By computing the ratio $ \delta \nu_\mathrm{fr} / \delta \nu_\mathrm{EC} $ using eq.~\ref{eq:intrinsic_lw_th} and considering how the EC configuration changes the emitted power and the mirror losses of the system, it is possible to compute the expected value for the internal losses (see section~D of Supplement~1). A value of $\alpha_{\mathrm{i}} = \SI{5.2}{cm^{-1}}$ is obtained. This value is in accordance with what expected for devices of the considered type~\cite{faist_quantum_2013}. 

\begin{figure}[htb!]
\centering\includegraphics[width=0.5\columnwidth, keepaspectratio]{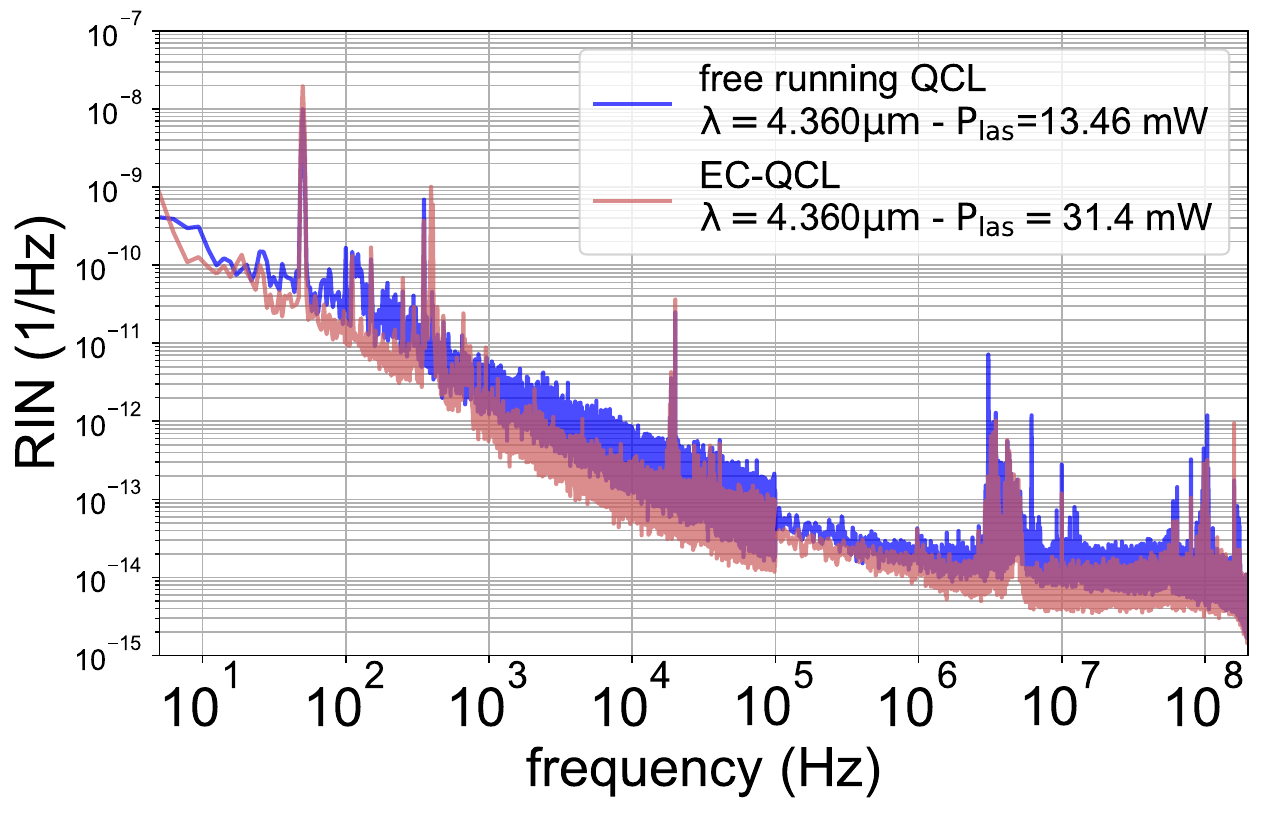}
\caption{Relative intensity noise of free-running (blue trace) vs EC-QCL operated at \SI{4.36}{\micro m} (red trace). The RIN is slightly reduced in the EC-QCL, proving that the grating does not induce additional intensity noise, except for the few peaks in the low frequency range caused by the grating's vibrations.}
\label{fig:RIN}
\end{figure}
Regarding the relative intensity noise (RIN), the spectra reveal that the RIN of the EC-QCL is comparable to or slightly lower than the RIN of the free-running QCL at any Fourier frequency, except for a few peaks in the 50~Hz--5~kHz caused by mechanical vibrations of the grating as visible in Fig.~\ref{fig:RIN}. We observe that the excess of noise due to the grating's vibrations is more pronounced in the FNPSD rather than in the RIN, showing that frequency fluctuations are more sensitive to mechanical vibrations than intensity fluctuations in QCLs. In general, the insensitivity of the RIN to optical feedback in QCLs, compared to standard laser diodes, had been demonstrated in ref.~\cite{zhao_relative_2019}. Plots of the RIN are available in Fig.~S5 of Supplement~1.
 
\section{Conclusions}
By implementing an external-cavity configuration based on a commercial diffraction grating, we have successfully induced a FP QCL to emit on a single frequency while ensuring a broad tunability of the wavelength. This very simple setup enhances the laser performance in terms of threshold current and emitted power. Our results demonstrates that the grating affects the laser's noise properties in a positive way, in particular, the intrinsic linewidth is substantially reduced and the full linewidth is significantly narrower for short integration times below 1 ms. The full linewidth is slightly narrower even for longer integration times, despite the residual peaks 
related to the mechanical vibrations of the grating. All of this comes together with a relative intensity noise (RIN) that is lower than or comparable to the RIN of the free-running QCL. In addition, by comparing the intrinsic linewidth measured in the two different configurations, the internal losses of the laser has been estimated, obtaining a value that is in agreement with what expected for this type of devices. The reduction of frequency and intensity noise is effective along the whole tuning range guaranteed by the grating. These characteristics make the EC-QCL a good candidate for spectroscopic applications such as absorption spectroscopy of $\mathrm{CO_2}$, which has strong absorption lines in the 4.20--\SI{4.35}{\micro m} range\cite{galli_absolute_2013}. 


\section*{Acknowledgments} 

The authors acknowledge financial support by the European Union's NextGenerationEU Programme with the I-PHOQS Infrastructure [IR0000016, ID D2B8D520, CUP B53C22001750006] "Integrated infrastructure initiative in Photonic and Quantum Sciences"; by the European Union's Research and Innovation Programme Horizon Europe with the Laserlab-Europe Project [G.A.~n.~871124] and the MUQUABIS Project [G.A.~n.~101070546] "Multiscale quantum bio-imaging and spectroscopy"; by the European Union's QuantERA II [G.A.~n.~101017733] -- QATACOMB Project "Quantum correlations in terahertz QCL combs"; by the Italian ESFRI Roadmap (Extreme Light Infrastructure -- ELI Project); by the Italian Ministero dell'Università e della Ricerca (project PRIN-2022KH2KMT QUAQK); by ASI and CNR under the Joint Project "Laboratori congiunti ASI-CNR nel settore delle Quantum Technologies (QASINO)" (Accordo Attuativo n. 2023-47-HH.0); by the European Partnership on Metrology with the 23FUN04 COMOMET project [Funder ID 10.13039/100019599].

\section*{Disclosures}
The authors declare no conflicts of interest.

\section*{Data Availability}
Data underlying the results presented in this paper are available from the corresponding author upon reasonable request.

\section*{Supplemental document}
See Supplement~1 for supporting content.



\printbibliography

\end{document}